\documentclass[onecolumn,12pt,preprintnumbers,amsmath,amssymb]{revtex4}

\usepackage{epsf}
\usepackage{graphicx}  
\usepackage{dcolumn}   
\usepackage{bm}        

\newcommand{\be}{\begin{equation}}
\newcommand{\en}{\end{equation}}
\newcommand{\ba}{\begin{array}}
\newcommand{\ea}{\end{array}}
\newcommand{\bea}{\begin{eqnarray}}
\newcommand{\nn}{\nonumber}
\newcommand{\eea}{\end{eqnarray}}

\begin{document}

\preprint{hep-th/0505186}

\title{Crossing $w=-1$ in Gauss-Bonnet Brane World \\
with Induced Gravity}

\author{Rong-Gen Cai\footnote{E-mail address: cairg@itp.ac.cn} and
  Hong-Sheng Zhang\footnote{E-mail address: zhanghs@itp.ac.cn}}
\affiliation{Institute of Theoretical Physics, Chinese Academy of
Sciences, P.O. Box 2735, Beijing 100080, China}

 \author{Anzhong Wang\footnote{E-mail address: anzhong$\_$wang@baylor.edu}}
\affiliation{CASPER, Department of Physics, Baylor University, 101
Bagby Avenue, Waco, TX 76706, USA}


\begin{abstract}
Recent type Ia supernovas data seemingly favor a dark energy model
whose equation of state $w(z)$ crosses $-1$ very recently, which
is a much more amazing problem than the acceleration of the
universe. In this paper we show that it is possible to realize
such a crossing without introducing any phantom component in a
Gauss-Bonnet brane world with induced gravity, where a four
dimensional curvature scalar on the brane and a five dimensional
Gauss-Bonnet term in the bulk are present. In this realization,
the Gauss-Bonnet term and the mass parameter in the bulk play a
crucial role.
\end{abstract}


 \maketitle

\section{Introduction}
 The existence of dark energy is one of the most significant cosmological discoveries
 over the last decade \cite{acce}. However, the nature of this dark energy remains a
 mystery. Various models of dark energy have been proposed, such as a small positive
 cosmological constant, quintessence, k-essence, phantom, holographic dark energy,
 etc., see \cite{review} for
 recent reviews with fairly complete lists of references of different dark energy
models. A cosmological constant is a  simple candidate for dark
  energy. However, following the more accurate data a more dramatic result
  appears:
  the recent analysis of the type Ia supernovas data
  indicates that the time varying dark energy gives a better
  fit  than a cosmological constant, and in particular, the equation of state parameter
   $w$ (defined as the ratio of
 pressure to energy density) crosses $-1$ at $z=0.2$ from above to below
 \cite{vari}, where $w=-1$ is the equation of state for the
 cosmological constant.  The dark energy with $w<-1$ is called phantom dark
 energy~\cite{call}, for which all energy conditions are
 violated.
  To obtain $w <-1$, scalar field with a negative kinetic term,
  may be a simplest realization \cite{phantom}. However,
 the equation of state of phantom scalar field is  always
 less than $-1$  and can not cross $-1$. Also it has been shown
 that the equation of state cannot cross $-1$ in the k-essence
 model of dark energy under some reasonable
 assumptions~\cite{Vik}.
  Some dark energy models which contain a negative-kinetic scalar field and a normal
scalar field  have been considered in \cite{guozk}; in these
models crossing the border $w=-1$ can be realized. More recently
it has been found that crossing $-1$ within a single scalar field
model is possible, if its action contains higher derivative
terms~\cite{Li} (see also \cite{Binflation}) .

In this paper we suggest a possibility with effective equation of
state crossing $-1$ in brane world scenario. The brane world
scenario is now one of the most important ideas in high energy
physics. In this scenario, the standard model particles are
confined on the 3-brane, while the gravitation can propagate in
the whole spacetime. As for cosmology in the brane world scenario,
many works have been done over the last several years; for a
review, see \cite{maarten} and references therein. Brane world
models admit a wider range of possibilities for dark
energy~\cite{sahni}. For example, the so-called DGP model has a
late-time self-acceleration solution~\cite{dgp}.  On the other
hand,
 higher derivative curvature terms, for example $R^2$ in Euler density,
   naturally appear in many occasions,
  such as in the
quantum field theory in curved space \cite{qftincur}. Generally
speaking,  resulting equations of motion of such terms contain
more than second derivatives of metric and the theory is plagued
by ghosts. However there exists a combination of quadratic terms,
called Gauss-Bonnet term, whose equation of motion has no more
than second derivatives of metric and the theory is free of ghosts
\cite{earlygb}.  Another important property of Gauss-Bonnet term
is that the Lagrangian is a pure divergence in four or less
dimensions. So only in more than four dimensional theories, the
Gauss-Bonnet combination has dynamically physical
 meanings. In addition, the Gauss-Bonnet term naturally appears in
 low energy effective action of  string theory
 \cite{stringgb1,stringgb2}. The study on the effects of
  the Gauss-Bonnet term in the bulk of brane world
 model is therefore well motivated.

 The gravitational action can be generalized in various ways in
  brane world scenario. A four dimensional scalar
curvature term on the brane is an important one except a
Gauss-Bonnet term in
 the bulk.  This induced gravity correction arises because the
localized matter fields on the brane, which couple to bulk
gravitons, can generate via quantum loops a localized
four-dimensional world-volume kinetic term for
gravitons~\cite{dgp}. The combined effect of these curvature
corrections to the action can remove the infinite-density big bang
singularity, although the Gauss-Bonnet correction, which is
expected to dominate at early times, on its own does not remove
the infinite-density singularity, while the induced gravity
correction on its own mostly affects the late-time evolution
\cite{combinecos}. We find that the combining effect of the
Gauss-Bonnet term in the bulk and the induced gravity term on the
brane behaves as dark energy on the brane, and the effective
equation of state can cross the phantom divide $w=-1$.

Here we would like to mention that a dark energy model where
standard gravity with dilaton scalar field  containing an
additional scalar-dependent coupling with Gauss-Bonnet invariant
has been investigated recently in \cite{nojiri}; The effect the
higher derivative terms on the big rip of phantom cosmology has
been discussed in \cite{Sami}.

\section{The Model}
 Let us start with the action
  \be
  \label{action}
  S=S_{\rm bulk}+S_{\rm brane},
  \en
 where
  \be
  S_{\rm bulk}=\frac{1}{16 \pi ^{(5)}G}\int d^5x \sqrt{{\rm det} (^{(5)}g)}
  \left(^{(5)}R-2\Lambda_5+\alpha L_{GB}\right),
  \en
  \be
  L_{GB}= ^{(5)\!}R^{2}
  -4\,^{(5)\!}R_{AB}\,^{(5)\!}
  R^{AB}+^{(5)\!}R_{ABCD}\,^{(5)\!}R^{ABCD}\,,
  \en
 and
   \be
  S_{\rm brane}=\frac{1}{16\pi ^{(4)}G} \int_{y=0}d^4x \sqrt{{\rm det} (^{(4)}g)}
  \left(^{(4)}R-2\Lambda_4\right).
  \en
  Here $S_{\rm bulk}$ is the action of the bulk, $S_{\rm brane}$ is
  the action of the brane, $S$ is the total action, $^{(5)}G$
  stands for the five dimensional Newton constant in the bulk,
  $^{(4)}G$ represents the four
  dimensional Newton constant on the brane, $\Lambda_5$ denotes cosmological
  constant in the bulk, $\Lambda_4$ is cosmological constant
  on the brane, $L_{GB}$ represents the Gauss-Bonnet term,
  and $\alpha$ is a constant with dimension $[length]^2$.
    $^{(4)}g$ is the induced metric on the brane
  \be
  ^{(4)}g=^{(5)}g-(n,n)^{-1} n\otimes n,
  \en
  where $n$ is the normal vector of the brane. We choose the coordinate of extra dimension
  $y$ such that the brane stands
  at $y=0$. Denoting
  \be
  r_c=\frac{^{(5)}G}{^{(4)}G},
  \en
 one can see that $r_c$ has dimension of $[length]$.

  Assuming there is a  mirror symmetry in the bulk, we
  have the Friedmann equation on the brane \cite{combinecos},
  see also \cite{cov},
  \bea
 {4\over r_c ^2}\left[1 +\frac{8}{3}\alpha\left(H^2 +{k \over a^2}
 + {U \over 2} \right) \right]^{2}\left(H^2 +{k \over
 a^2}-U\right) \nn
  \\ =\left(H^2 +{k \over a^2} -\frac{8\pi ^{(4)}G}
 {3}(\rho+\lambda)\right)^{2}\,,
 \label{fried}
 \eea
 where $H=\dot{a}/a$ is the Hubble parameter, $a$ is the
 scale factor, $k$ is the spatial curvature of the
  three dimensional maximally symmetric space in the FRW metric on the brane, and
 \be
 \lambda=\frac{\Lambda_4}{8\pi {}^{(4)}G},
 \en
 \be
 \label{U}
  U=-\frac{1}{4\alpha}\pm \frac{1}{4\alpha} \sqrt{1+
  4\alpha\left(\frac{\Lambda_5}{6}+\frac{2M{~}^{(5)}G}{a^4}\right)}.
  \en
  Here $M$ is a constant, standing for the mass of bulk black
  hole. Note that when one takes the positive sign, the above
equation can be reduced to the case of generalized DGP
model~\cite{dgp}(In the pure DGP model, $\Lambda_4=\Lambda_5=0$).
In the latter case, we have
  \be
  H^2+\frac{k}{a^2}=\frac{8\pi ^{(4)}G}{3}(\rho+\lambda)+
  \frac{2}{r_c^2}\pm \frac{1}{\sqrt{3}r_c} \left[ 4\left(8\pi
  ^{(4)}G\right)^2 (\rho+\lambda)
  -2\Lambda_5+\frac{12}{r_c^2}-\frac{24M
  ^{(5)}G}{a^4}\right]^{1/2}.
 \label{dgp}
  \en
  Its various limits and inflation models have been studied in \cite{PZ}.
Note that the branch with minus sign in (\ref{U}) cannot be
reduced to the case of the generalized DGP model as $\alpha \to
0$. Therefore we only consider the branch with the plus sign in
(\ref{U}) in what follows.

  Introduce the
 following new variables and parameters,
 \begin{eqnarray}
 \label{parameters}
 && x \equiv \frac{H^2}{H_0^2}+\frac{k}{a^2H_0^2}=
   \frac{H^2}{H_0^2}-\Omega_{k0} (1+z)^2,  \nonumber \\
 && u \equiv \frac{8\pi \ ^{(4)}G}{3H_0^2}(\rho +\lambda)=
 \Omega_{m0}(1+z)^3+\Omega_{\lambda}, \nonumber \\
 && m \equiv \frac{8}{3} \alpha H_0^2, \nonumber
\\
&&  n \equiv \frac{1}{H_0^2r_c^2}, \nonumber \\
 &&
 y  \equiv \frac{1}{2}UH_0^{-2}=\frac{1}{3m}\left(-1+\sqrt{1+
 \frac{4\alpha \Lambda_5}{6}+{\frac{8\alpha M
 {}^{(5)}G}{a^4}}}\right) \nonumber \\
 &&~~~=
 \frac{1}{3m}\left(-1+\sqrt{1+m\Omega_{\Lambda_5}+m\Omega_{M0}(1+z)^4}\right),
 \end{eqnarray}
 where $H_0$ is the present value of Hubble parameter, $z$ denotes the red shift with definition
 $z= a_0/a-1$, and we have assumed that there is only pressureless dust in the universe. In addition,
 we have used the following notations
 \begin{equation}
 \label{Omega}
\Omega_{k0}=-\frac{k}{a_0^2H_0^2},~~\Omega_{m0}=\frac{8\pi
  ^{(4)}G}{3}\frac{\rho_{m0}}{H_0^2},~~\Omega_{\lambda}=\frac{8\pi
  ^{(4)}G}{3}\frac{\lambda}{H_0^2},
  ~~\Omega_{\Lambda_5}=
  \frac{3\Lambda_5}{8H_0^2},~~\Omega_{M0}=\frac{3M
  ^{(5)}G}{a_0^4H_0^2}.
  \end{equation}
With these new
 variables and parameters,
 (\ref{fried}) can be rewritten as
 \be
 \label{GBI}
 4n(x-2y)[1+m(x+y)]^2=(x-u)^2.
 \en
 This is a cubic equation of the variable $x$. According to algebraic theory it
 has 3 roots. One can explicitly write down three roots. But they are too lengthy and
 complicated to present here. Instead we only express those three roots formally
 in the order given in {\it Mathematica}
 \begin{eqnarray}
 \label{solution}
 && x_1= x_1(y,u|m,n), \nonumber \\
 && x_2=x_2(y,u|m,n), \nonumber \\
&& x_3=x_3(y,u|m,n),
\end{eqnarray}
where $y$ and $u$ are two variables, $m$ and $n$ stand for two
parameters. The root on $x$ of the equation (\ref{GBI}) gives us
the modified Friedmann equation on the Gauss-Bonnet brane world
with induced gravity. From the solutions given in
(\ref{solution}), this model seems to have three branches.
 In addition,
note that all parameters introduced in (\ref{parameters}) and
(\ref{Omega}) are not independent of each other. According to the
Friedmann equation (\ref{solution}), when all variables are taken
current values, for example, $z=0$, the Friedmann equation will
give us a constraint on those parameters
\begin{equation}
\label{constraint}
1=f(\Omega_{k0},~\Omega_{m0},~\Omega_{M0},~\Omega_{\Lambda_5},
  ~\Omega_{\lambda},~m,~n).
\end{equation}

 Now we turn to the evolution of the universe in the brane world scenario, and pay particular attention
 to the acceleration characteristic of expansion at low red shift.
 To some issues on the dark energy, one of the most significant parameters from the viewpoint of
  observations is the equation of state,  $w={pressure}
  /{energy~density}$, of the dark energy. In general relativity
  the dark energy has to be introduced in order to coincide with
  observation data. In our model, the effect of the bulk
  Gauss-Bonnet term and induced gravity term on the brane may play the role of dark energy
  in general relativity. Therefore, in our model, the accelerated expansion
  of the universe is due to the combined effect of the brane world scenario and
  the Gauss-Bonnet term in the bulk.

 To explain observed accelerated expansion, we calculate the equation of state $w$ of the effective
  ``dark energy" caused by the bulk Gauss-Bonnet term and induced gravity term
  by comparing the modified Friedmann equation in
  the brane world scenario and the standard Friedmann equation in general
  relativity, because all observed features of dark energy are
  ``derived" in general relativity.
  Note that the Friedmann equation in the
four dimensional
  general relativity can be written as
 \be
 H^2+\frac{k}{a^2}=\frac{8\pi ^{(4)}G}{3} (\rho+\rho_{de}),
 \label{genericF}
 \en
 where the first term of RHS of the above equation represents the dust matter and the second
 term stands for the dark energy. Suppose that dark energy itself satisfies
  the continuity equation
 \be
 \frac{d\rho_{de}}{dt}+3H(\rho_{de}+p_{de})=0,
 \label{em}
 \en
 we can then express the equation of state for the dark
 energy as
   \be
  w_{de}=\frac{p_{de}}{\rho_{de}}=-1-\frac{a}{3}\frac{d \ln \rho_{de}}{da}.
   \en
 Note that we can rewrite the Friedmann equation (\ref{solution}) in the form of
 (\ref{genericF}) as
 \be
 \label{eq19}
 xH_0^2=\frac{8\pi ^{(4)}G}{3} \rho+\left(H_0^2 x(y,u|m,n)-\frac{8\pi ^{(4)}G}{3}
 \rho\right)=\frac{8\pi ^{(4)}G}{3}( \rho+F),
 \en
 where $\rho$ is the energy density of dust matter on the brane
 and the term
 $$F \equiv \left(\frac{3H_0^2}{8\pi ^{(4)}G} x(y,u|m,n)-
 \rho\right)$$
 corresponds to $\rho_{de}$ in (\ref{genericF}).  In addition, here $x$ in (\ref{eq19}) denotes one
 of three solutions given in (\ref{solution}).  Now we  regard $F$ in (\ref{eq19})
 as the effective dark energy density,  then we have the equation of state for the
 effective dark energy
 \be
 \label{effw}
 w_{eff}=-1-\frac{a}{3}\frac{d\ln F}{da}=-1+\frac{1}{3}\frac{d\ln F}{d\ln (1+z)}.
 \en
The type Ia supernovas observation data indicate that the equation
of state of dark energy undergoes a transition from $w>-1$ to
$w<-1$ at $z=0.2$~\cite{vari}. As pointed out above, it is quite
difficult to realize such a transition in usual field theory
models. Here we will see that it can be realized in our
Gauss-Bonnet brane world scenario with induced gravity, in which
the Gauss-Bonnet term and the mass parameter $M$ in bulk play an
important role.

Before proceeding, let us first discuss the case without the
Gauss-Bonnet term, namely $\alpha =0$.  In this case, as can be
seen from (\ref{dgp}), the effective dark energy density
 is
  \be
  \label{FDGP}
  F=\frac{3}{8 \pi ^{(4)}G} \left\{\frac{2}{r_c^2}\pm \frac{1}{\sqrt{3}r_c} \left[ 4\left(8\pi
  ^{(4)}G\right)^2 (\rho+\lambda)
  -2\Lambda_5+\frac{12}{r_c^2}-\frac{24M
  ^{(5)}G}{a^4}\right]^{1/2}\right\}.
  \en
  When the bulk mass parameter $M$ vanishes, namely, $M=0$, this form of
  effective dark energy has been
  discussed in \cite{sahni}. For the branch with plus sign in (\ref{FDGP})
   ( called branch 2 in \cite{sahni}), the equation of state for the effective dark
  energy is always larger than $-1$, and approaches to $-1$ as
  $z \to -1$.  On the other hand, for the branch with minus sign
  in (\ref{FDGP}) ( called branch 1 in \cite{sahni}), there is a
  divergence on the curve of $w_{eff}$ versus the red shift $z$
  at some point $z_0$, and in the region $z\in (-1,z_0)$, $w_{eff}
  $ monotonically goes from $-\infty$ at $z=z_0$ to $-1$ at $z=-1$,
  while in the region $z \in (z_0, \infty)$,  it monotonically
  goes from $+\infty$ at $z_0$ to some value larger than $-1$ at
  $z=\infty$. Therefore in the branch 1, the equation of state
  for the effective dark energy can be less than $-1$, but
  it can never cross the phantom divide $-1$.  In addition, let
  us remind that the divergence of the equation of state at some
  point $z=z_0$, whose value depends on the model parameters,
  is harmless. This divergence is caused by $F=0$ there,  except
  for the equation of state $w_{eff}$, other physical quantities
  like Hubble parameter, deceleration factor, energy density of the effective dark
  energy, etc., behave well.

  When the bulk mass parameter $M$ does not vanish, namely $M \ne
  0$, one may ask whether this situation will be changed. The
  answer is no.  In this case, the equations of state for two branches are
  \begin{eqnarray}
  \label{dw}
  && w^+=-1+\frac{3Ia-8J}{6B(a^2+B)}, \nonumber \\
  && w^-=-1-\frac{3Ia-8J}{6B(a^2-B)},
  \end{eqnarray}
  where $w^+$ and $w^-$ stands for the $``+"$ and $``-"$ branch in
  (\ref{FDGP}), respectively, and
  $$ I = \rho_{m0} a_0^3,\ \ \ B=\sqrt{a^4+aI -2J'}, \ \ \ J= A r_c^{-2} \ ^{(5)}G.$$
Note that $w^+$ diverges at $B=0$, and $w^-$ diverges at $a^2 =B$
or $B=0$.  We further find from (\ref{dw}) that if $B\neq 0$, we
have $w^+=-1$ and $w^-=-1$ when
$$ a= \frac{8J}{3I}.$$
However, we can show that
$$ \left.\frac{dw^+}{da}\right|_{a=\frac{8J}{3I}} >0, \ \ \
\ \ \left.\frac{dw^-}{da}\right|_{a=\frac{8J}{3I}} >0. $$
 Both cases are not consistent with the observation fact: the
 equation of state of dark energy transits at $z=0.2$ from $w>-1$ for $z>0.2$ to $w<-1$
 for $z<0.2$~\cite{vari}.

  Next we turn to the case with $\alpha \neq 0$, namely the case with a
  Gauss-Bonnet term in the bulk. For simplicity, we set
  $k=0$ and $\Lambda_4=\Lambda_5=0$ in what follows. To see the
  properties of three branches in (\ref{solution}), let us first look
  at the asymptotical behavior of them as $z \to -1$. In this
  case, the equation (\ref{GBI}) reduces to
  \begin{equation}
  \label{eq23}
  4n x (1+mx)^2=x^2,
  \end{equation}
  which has three roots
  \begin{eqnarray}
  \label{eq24}
  && \tilde x_1=0, \nonumber \\
  && \tilde x_2= \frac{1-8nm+\sqrt{1-16nm}}{8nm^2}, \nonumber\\
  && \tilde x_3=\frac{1-8nm-\sqrt{1-16nm}}{8nm^2}.
  \end{eqnarray}
  We find that when $nm>1/16$, three solutions in (\ref{solution})
  go to the limits in (\ref{eq24}) in the following manners
  \begin{equation}
  \label{eq25}
  x_1 \to \tilde x_1, \ \ \ x_2 \to \tilde x_2,\ \ \ x_3 \to
  \tilde x_3;
  \end{equation}
  when $nm<1/16$,
  \begin{equation}
  \label{eq26}
  x_1 \to \tilde x_2,\ \ \ x_2 \to \tilde x_1,  \ \ \ x_3
  \to \tilde x_3;
  \end{equation}
  and when $nm=1/16$,
  \begin{equation}
  \label{eq27}
x_1 \to 1/m, \ \ \ x_2 \to 0, \ \ \ x_3 \to 1/m.
\end{equation}
From the above, we can see that (1) when $nm>1/16$, only does the
solution $x_1$ make sense, while $x_2$ and $x_3$ are not physical
solutions since $\tilde x_2$ and $\tilde x_3$ are imaginary; (2)
when $nm<1/16$, three roots $\tilde x_1$, $\tilde x_2$ and $\tilde
x_3$ in (\ref{eq24}) look reasonable since they are all real. But
we note that when $m\to 0$, the branches $x_2$ and $x_3$ reduce to
two branches in the DGP model, while $x_1 \to \infty $. So $x_1$
is a new branch in our model and has no corresponding one in the
DGP model. The two branches in the DGP model also can be obtained
directly from (\ref{solution}):
\begin{eqnarray}
\label{eq28}
 && \lim_{m \to 0}x_2 = u + 2n -2 \sqrt{n^2+nu}, \nonumber \\
 && \lim_{m\to 0}x_3 = u + 2n +2 \sqrt{n^2+nu}.
 \end{eqnarray}

 We now consider the case $M=0$ in our model. Assuming
 $\Omega_{m0}=0.28$, which is indicated by WMAP data~\cite{WMAP}, we find that
 for any values of $m$ and $n$, the condition
\begin{equation}
\label{eq29}
 x|_{z=0}=1,
 \end{equation}
 always cannot be satisfied for the branches $x_1$ and $x_2$ in (\ref{solution}),
 while $x_3$ always becomes imaginary at finite red shift.
 Therefore the case with a vanishing bulk mass, $M=0$, cannot
 satisfy our requirement. In Fig.~1 we plot the $x$ for the
 branch $x_1$ with respect to the parameters  $m$ and $n$ at red shift $z=0$.
\begin{figure}
 \centering
 \includegraphics[totalheight=3in]{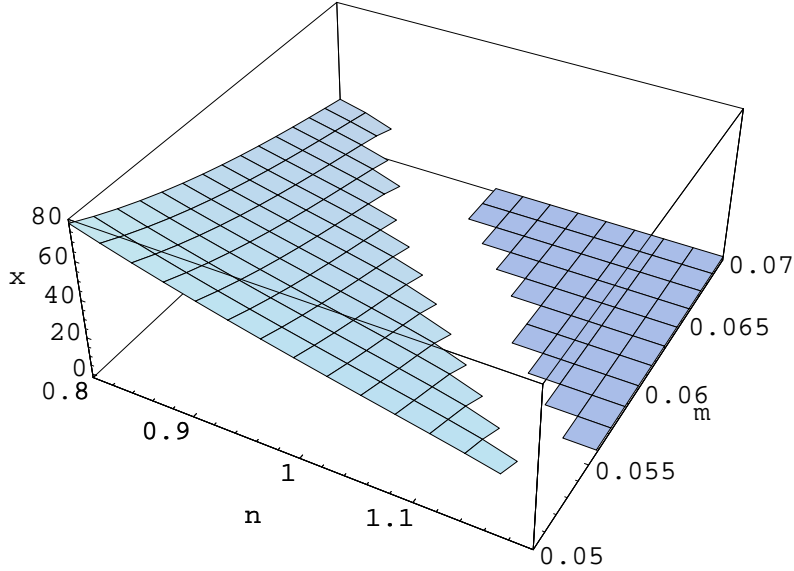}
 \caption{$x_1|_{z=0}$  with $M=0$ and $\Omega_{m0}=0.28$ versus the parameters
 $m$ and $n$.
 It is obvious that $x_1(m,n)$ is not a continuous function. The discontinuity
 happens at $nm=1/16$. In the left
 part of the surface $x(m,n) \gg 1$, the right part $x(m,n)\ll 1$. There
 does not exist a point on which $x_1=1$. }
 \label{strange}
 \end{figure}

 The case with $M \ne 0$:

  (i) For the branch $x_1$ in (\ref{solution})
  , we find that it is possible to satisfy
  the condition $x_1|_{z=0}=1$ with $\Omega_{m0}=0.28$.  From the viewpoint of physics
  consistency we may think from (\ref{parameters}) that $m$ and
  $n$ should be of order one or less. And from (\ref{constraint})
  one can obtain
  \be
     \Omega_{M0}=\Omega_{M0}(m,n),
  \en
  with condition $x_1|_{z=0}=1$ and $\Omega_{m0}=0.28$. In Fig.~2
  we plot $\Omega_{M0}$ as a function of $m$ and $n$, which
  indicates that $\Omega_{M0}$ is positive and is of order one, as
  expected.
  \begin{figure}
 \centering
 \includegraphics[totalheight=3in]{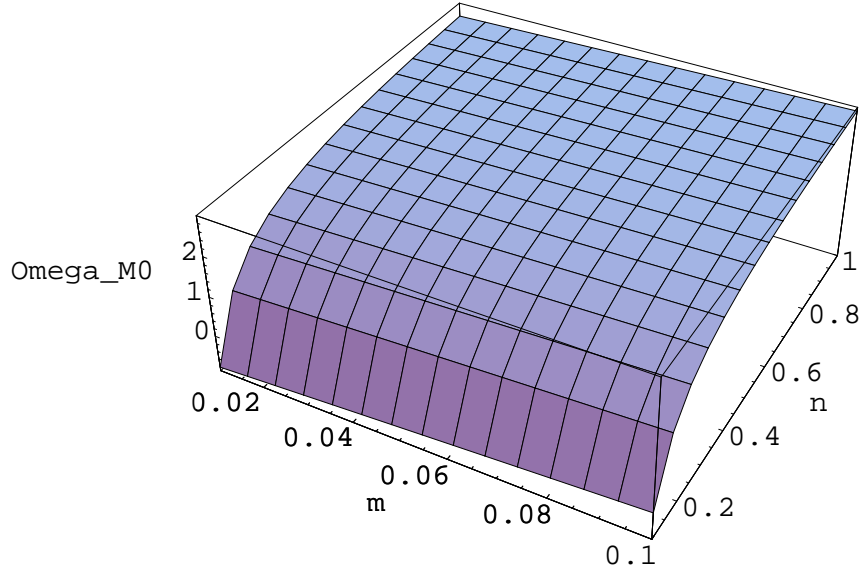}
 \caption{$\Omega_{M0}$ versus $m$ and $n$,  with
 $\Omega_{m0}=0.28$ and $k=0$.}
 \label{omegaa}
 \end{figure}

 In Fig.~3 we show the equation of state for the effective dark
 energy when we take $m=1.036$ and $n=0.04917$. In this case, from
 the constraint equation (\ref{constraint}), one has
 $\Omega_{M0}=2.08$. From the figure we see that $w_{eff}<-1$ at
 $z=0$ and
 $$ \left. \frac{dw_{eff}}{dz}\right|_{z=0}<0.
 $$
Therefore the equation of state for the effective dark energy can
indeed cross the phantom divide $w=-1$ near $z \sim 0$.

\begin{figure}
 \centering
 \includegraphics[totalheight=2.5in]{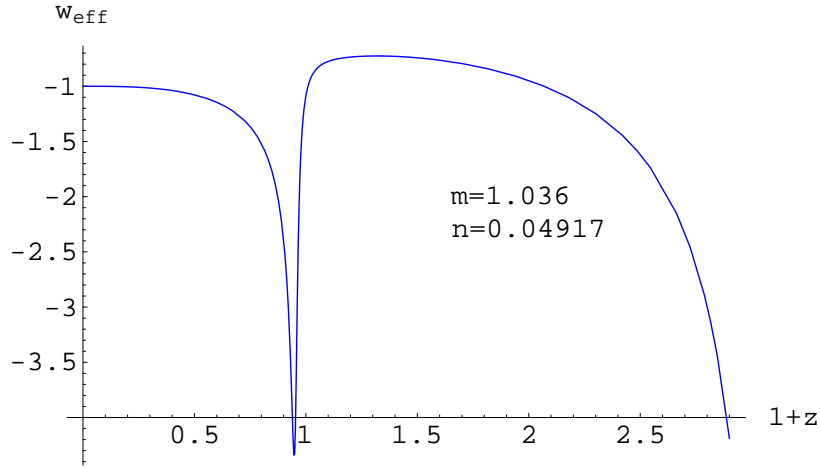}
 \caption{The equation of state $w_{eff}$ with respect to the
 red shift $1+z$, with $\Omega_{m0}=0.28$ and  $\Omega_A=2.08$.}
 \label{wm03}
 \end{figure}
Some remarks are in order. First when $nm>1/16$, one can also
realize  crossing $w=-1$ for the equation of state in this model.
However, in this case, one may find that the effective dark energy
density $F$ in (\ref{eq19}) is always negative, which cannot drive
the universe to accelerated expand. Therefore the stability
condition $nm<1/16$ has to be satisfied. Second, as the case of
generalized DGP model, the equation of state $w_{eff}$ in this
model has also a divergence at the some red shift point, which
depends on the model parameters. The divergence is  harmless and
it is also caused by $F=0$ at that point. To see the harmlessness,
In Fig.~4 and 5 we  respectively plot the Hubble parameter and the
fraction energy density of dust matter in the case of $m=1.036$
and $n=0.04917$. From Fig.~4 we can see that when $z<0$, the
Hubble parameter is an increasing function of red shift, which
shows the phantom behavior of dark energy. The universe finally
approaches to a de Sitter space.
\begin{figure}
 \centering
 \includegraphics[totalheight=2.5in]{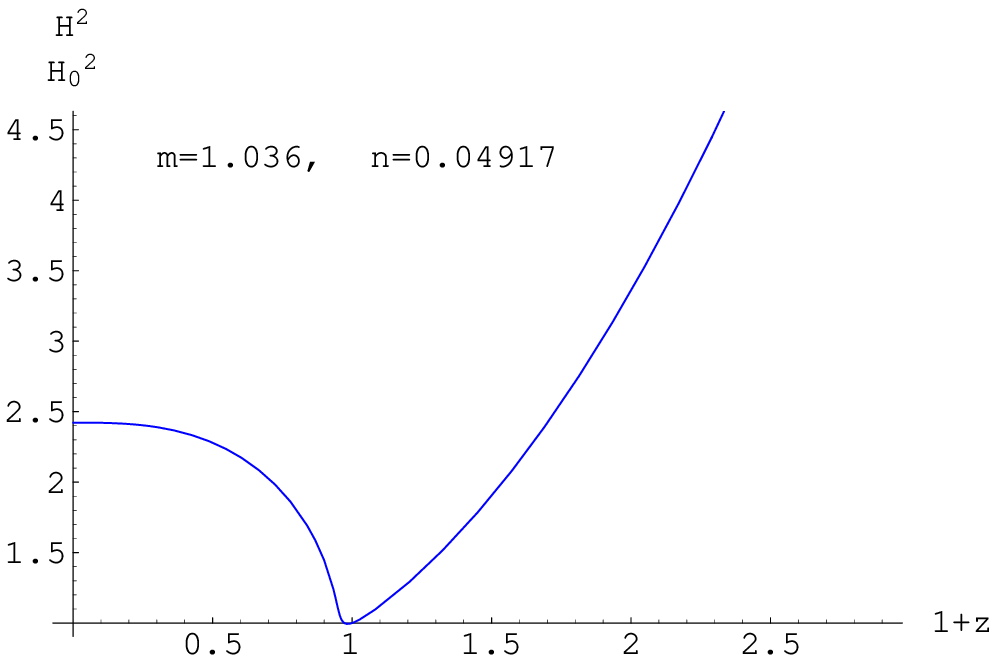}
 \caption{$H^2/H_0^2$ versus $1+z$,  with $\Omega_{m0}=0.28$
 and $\Omega_A=2.08$.}
 \label{hm03}
 \end{figure}

 \begin{figure}
 \centering
 \includegraphics[totalheight=2.5in]{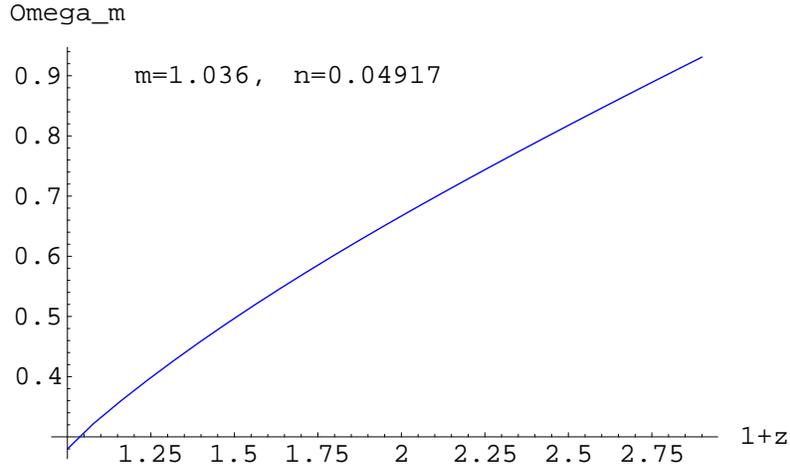}
 \caption{$\Omega_m$ versus $1+z$,  with $\Omega_{m0}=0.28$
 and $\Omega_A=2.08$.}
 \label{prom03}
 \end{figure}
To further show  the harmlessness of the divergence in the
equation of state $w_{eff}$, we would like to plot in Fig.~6 the
deceleration parameter $q$, which is defined by
 \be
  q=-\frac{\ddot{a}a}{\dot{a}^2}=\frac{1}{2}[\Omega_m+
  (1-\Omega_m)(1+3w_{eff})].
  \en
\begin{figure}
 \centering
 \includegraphics[totalheight=2.5in]{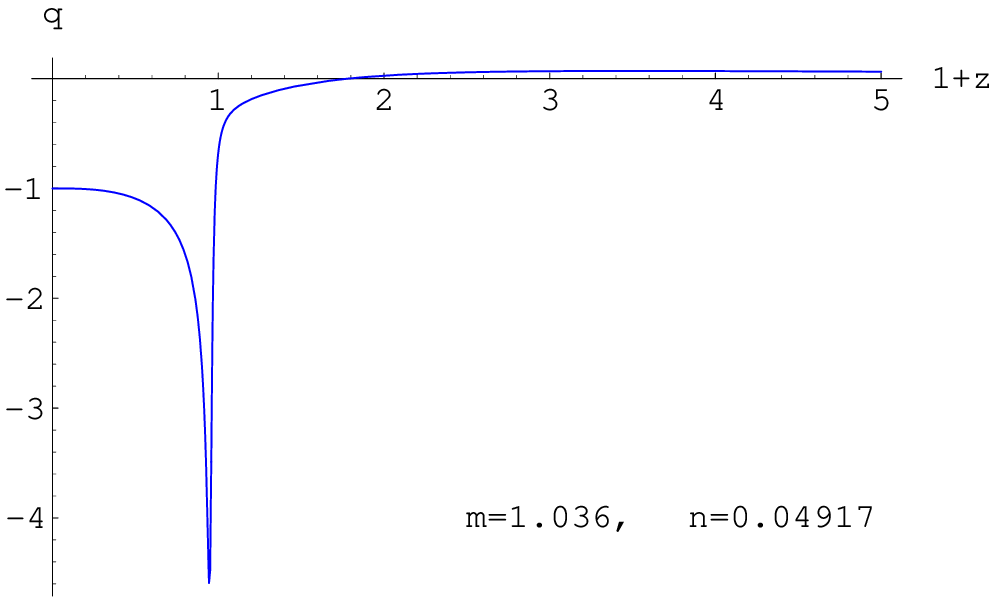}
 \caption{$q$ versus $1+z$, with $\Omega_{m0}=0.28$ and
 $\Omega_A=2.08$.}
 \end{figure}
 The transition of the universe from deceleration to acceleration phase
 happens at $z=0.8$ in this parameterized model.

 In the above demonstration, we take $\Omega_{m0}=0.28$. In
 Fig.~7,8 and 9, we further plot the corresponding quantities for the case
 with $\Omega_{m0}=0.05$, $m=1.17$ and $n=0.0511$.

  \begin{figure}
 \centering
 \includegraphics[totalheight=2.5in]{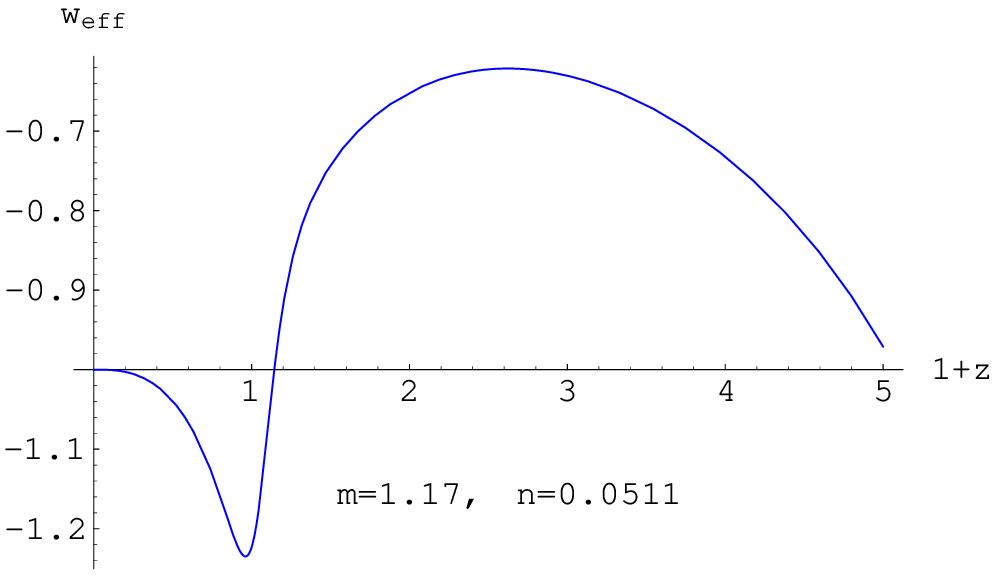}
 \caption{$w_{eff}$ versus $1+z$, with $\Omega_{m0}=0.05$ and
 $\Omega_A=0.407$.}
 \end{figure}

 \begin{figure}
 \centering
 \includegraphics[totalheight=2.5in]{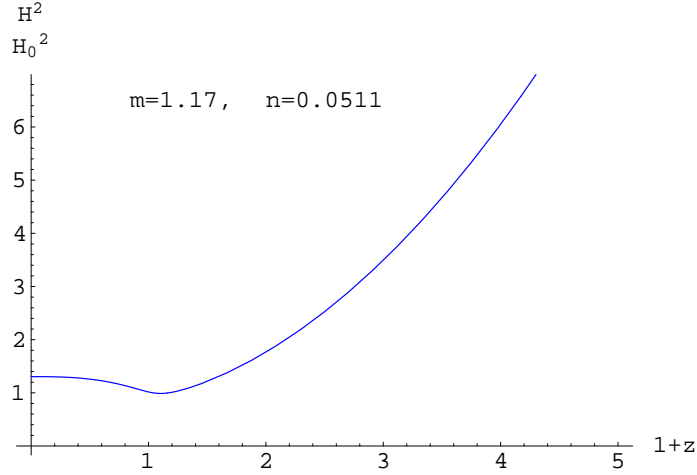}
 \caption{$H^2/H_0^2$ versus $1+z$,  with $\Omega_{m0}=0.05$
 and $\Omega_A=0.407$.}
 \end{figure}

 \begin{figure}
 \centering
 \includegraphics[totalheight=2.5in]{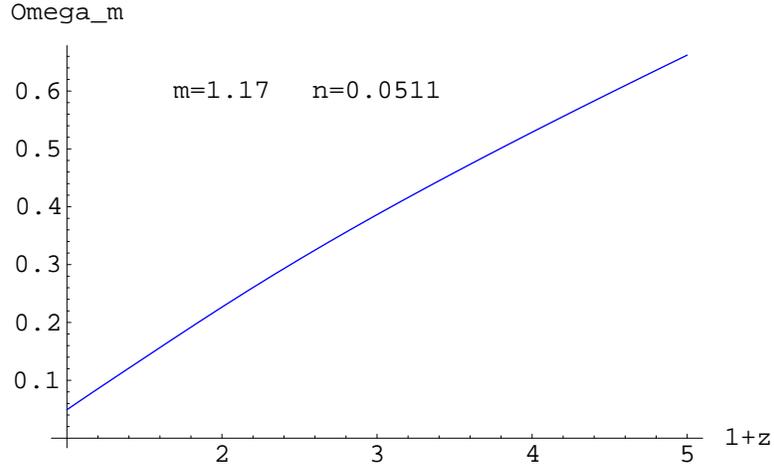}
 \caption{$\Omega_m$ versus $1+z$,  with $\Omega_{m0}=0.05$
 and $\Omega_A=0.407$.}
 \end{figure}

\begin{figure}
 \centering
 \includegraphics[totalheight=2.5in]{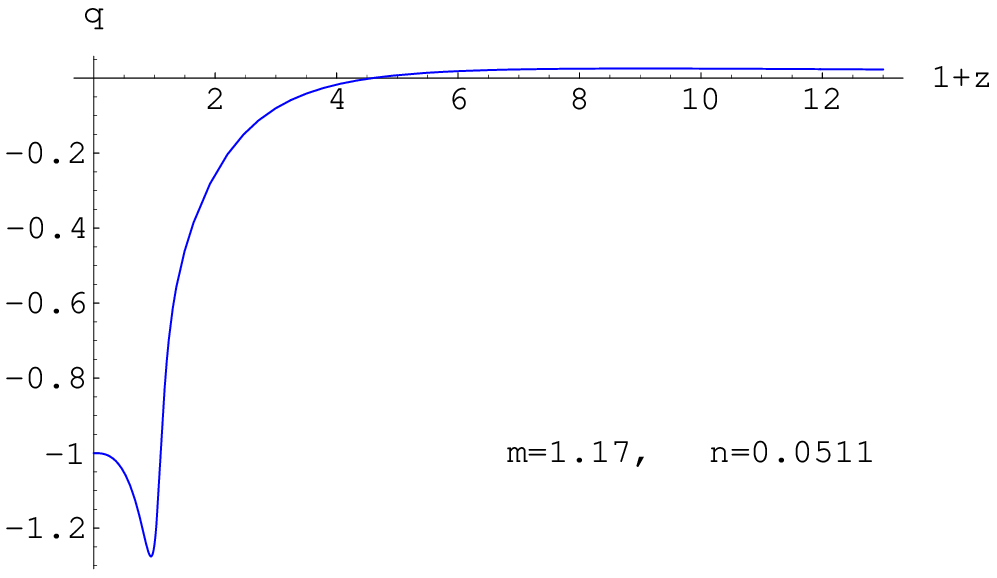}
 \caption{$q$ versus $1+z$, with $\Omega_{m0}=0.05$ and
 $\Omega_{M0}=0.407$.}
 \end{figure}

(ii) For the branches $x_2$ and $x_3$ in (\ref{solution}), we find
that they always become imaginary at some finite red shift. It can
be understood as follows. These two branches go to corresponding
ones of the generalized DGP model as $m \to 0$, respectively. For
the generalized DGP model, we can see from (\ref{FDGP}) that there
is a square root in the effective dark energy density and the
square root will always become imaginary as $a$ decreases if $M\ne
0$.

\section{Conclusion}

 To summarize, in this paper we have shown that in the Gauss-Bonnet brane world with
 induced gravity, where a Gauss-Bonnet term in the bulk and a scalar curvature term
 on the brane appear, it is possible to realize crossing the phantom divide $w=-1$ for
 the equation of state of the effective ``dark energy", without introducing any
 phantom component. The dark energy model can be consistent with astronomical observation
 data. In realizing the transition from $w>-1$ to $w<-1$, the
 Gauss-Bonnet term and the bulk mass parameter play a crucial
 role. Finally we point out that the qualitative conclusion will
 not be changed if we include non-vanishing cosmological constants on the brane and in
 the bulk, namely,  $\Lambda_4 \ne 0$ and $\Lambda_5\ne 0$. In addition, it would be of great
 interest to fix the parameters in this dark energy model, according to current type Ia
supernovas data. This issue is currently under investigation.


{\bf Acknowledgments:} R.G. Cai thanks C.G. Huang H. Wei, X.M.
Zhang and Z.H. Zhu for helpful discussions. This work was
supported in part by a grant from Chinese academy of sciences,
grants No. 10325525 and No.90403029 from NSFC, and by the ministry
of science and technology of China under grant No. TG1999075401.

\end{document}